\def\lsim{\mathrel{\scriptstyle{\buildrel < \over \sim}}}
\magnification 1200
\baselineskip=17pt

\centerline{\bf ON THE DECOUPLING OF LAYERED SUPERCONDUCTING}
\bigskip
\centerline{\bf FILMS IN PARALLEL MAGNETIC FIELD}
\vskip 50pt
\centerline{J. P. Rodriguez}
\medskip
\centerline{\it Theoretical Division,
Los Alamos National Laboratory,
Los Alamos, NM 87545 and}
\centerline{{\it Dept. of Physics and Astronomy,
California State University,
Los Angeles, CA 90032.}\footnote*{Permanent address.}}
\vskip 30pt
\centerline  {\bf  Abstract}
\vskip 8pt\noindent
The issue of the decoupling  of extreme type-II superconducting
thin films ($\lambda_L\rightarrow\infty$)
with weakly Josephson-coupled layers in magnetic
field parallel to the layers 
is considered via the corresponding frustrated $XY$
model used to describe the 
mixed phase in the critical regime.  For the general case 
of arbitrary field orientations 
such that the perpendicular magnetic field component 
is larger than the decoupling
cross-over scale characteristic of layered superconductors,
we obtain independent
parallel and perpendicular vortex lattices. 
Specializing to
the double-layer case, we compute the  parallel lower-critical
field with entropic effects included, and find that it vanishes
exponentially as temperature approaches the layer
decoupling transition in zero-field.  The parallel
reversible magnetization is also calculated in this
case, where we
find that it shows a cross-over phenomenon
as a function of parallel field in the intermediate
regime of the mixed phase in lieu of a true  layer-decoupling
transition.  
It is argued that such is the case for any finite
number of layers, since the isolated double layer represents
the weakest link.

\bigskip
\noindent
PACS Indices:  74.60.Ec, 74.20.De, 11.10.Lm, 75.10.Hk

\vfill\eject
\centerline{\bf I. Introduction}
\bigskip

The study of layered superconductivity has been reinvigorated by the
discovery of the  high-temperature oxide superconductors.$^1$  In
the case of the Bismuth (Bi) and Thallium (Tl) based compounds,
for example, the superconducting coherence length $\xi_c$ perpendicular
to the conducting planes is less than the separation between layers,
and London theory fails.$^1$  
Hence a Lawrence-Doniach (LD) type of description in terms of weakly
Josephson-coupled layers becomes necessary.$^2$
The following question then naturally arises:
does a layer decoupling transition,
aided perhaps by the introduction of a parallel magnetic field,$^3$
occur in such systems in addition to or in
place of the conventional 
type-II superconducting ones marked by the lower and
upper critical fields,$^4$ $H_{c1}(T)$ and $H_{c2}(T)$?

In the absence of external magnetic field parallel
to the layers,  a layer decoupling transition
is indeed predicted to exist theoretically at a critical
temperature $T_*$
that lies above that of
the intra-layer Kosterlitz-Thouless (KT) transition temperature,
$T_c$.$^{5-8}$  This result is  based on studies of the $XY$ model
with weakly coupled layers, which accurately describes
a layered superconductor in the absence of
fluctuations of the magnetic field;$^{9,10}$ e.g., in the 
intermediate regime of the mixed phase
found in extreme type-II superconductors, where 
it is appropriate to take the
limit $\lambda_L\rightarrow\infty$ for the London
penetration length $\lambda_L$.
Also, recent experiments on Bi-based high-temperature superconductors
find evidence for a superconducting transition at $T_c^c$, where
the c-axis resistivity vanishes, that lies a few tenths of a degree
Kelvin above a second superconducting transition temperature
$T_c^{ab}$, where the resistivity in the $ab$-planes vanishes.$^{11,12}$
Hence both theory and experiment find an extraordinary regime in
temperature, $T_c = T_c^{ab} < T < T_* = T_c^c$, where Josephson
tunneling between layers exists while the layers themselves
are {\it resistive} and show
no intra-layer phase coherence.  
(It has recently been argued in ref. 13 that
$T_c^c$ in fact marks a sharp crossover for the bulk
case.)   Last, it is worth pointing out that Monte-Carlo simulations
of the three-dimensional (3D) $XY$ model obtain similar
behavior in the presence of a large magnetic field perpendicular
to the layers.$^{10}$

The nature of layer decoupling in the presence of magnetic
field parallel to the layers, 
on the other hand, is less well understood.  Early work by
Efetov suggests that superconducting  layers decouple   at high 
parallel fields
$B_{\parallel} > B_*^{\parallel}(0) \sim \Phi_0/d^2\gamma$,
where $\Phi_0$ denotes the flux quantum,
$\gamma = (m_c/m_{ab})^{1/2}$ is
the mass anisotropy parameter, and $d$ denotes the separation
between layers.$^3$   The latter calculation
is based on a high-temperature series expansion
analysis of the LD model.
A recent study
of the same model  by Korshunov and Larkin that  
employs a Coulomb gas representation, however,
finds no such layer decoupling in the high-field limit for temperatures 
below the decoupling transition.$^{14}$
In this   scenario, therefore, the layers remain effectively
Josephson coupled up to the parallel upper-critical
field, $H_{c2}^{\parallel}(T)$.

In this paper, we shall also examine the decoupling of layered
superconductors in parallel magnetic field, but
in film geometries of thickness much less than 
the in-plane London penetration length, and at temperatures near
the zero-field decoupling transition at $T_*$.$^{5-7}$
Specifically, we consider thin films of
extreme type-II layered superconductors ($\lambda_L\rightarrow\infty$)
in the intermediate regime ($H_{c1} \ll H
\ll H_{c2}$) of the mixed phase, which can be described by a frustrated
$XY$ model with a finite number of $N$ weakly coupled layers.$^{10}$
By working with the Villain form of the latter,$^{7,15}$ we obtain first
that the thermodynamics factorizes  into independent
perpendicular and parallel parts
in the presence of magnetic field at arbitrary orientation
so long as the perpendicular component, $H_{\perp}$, 
of the latter is larger than the
Glazman-Koshelev decoupling cross-over scale,$^{16}$
$B_*^{\perp}\sim\Phi_0/d^2\gamma^2$.
The perpendicular thermodynamics is characterized
by the melting of
two-dimensional (2D) vortices$^{17}$ that are decoupled from
the parallel Josephson vortices, as well as
from the perpendicular 2D vortices
in adjacent layers.
This is a result of the fact that well-formed vortex loops traversing
a few or more layers within the film are absent in the present
limit of weak inter-layer coupling.$^7$
In particular,
the parallel Josephson vortices are unable to
make connections between perpendicular 2D vortices
in the same or in  adjacent layer if the nearest-neighbor
spacings between these perpendicular 2D
vortices is  much less  than the zero-temperature Josephson penetration
length, $\lambda_J(0)\sim\gamma d$. 
This situation  occurs precisely for perpendicular fields
that satisfy
$H_{\perp}\gg B_*^{\perp}$, as stipulated above.
Also, since we first take the limit of extreme type-II superconductivity,
we then have the inequality $B_*^{\perp} \gg H_{c1}^{\perp}\sim
\Phi_0/\lambda_L^2$.
This  means that the former requirement guarantees
that the distance between perpendicular
vortices is within the London penetration length, which
in turn guarantees that
magnetic screening transverse to the perpendicular field component
is negligible.
The parallel thermodynamics, on the other hand, is
described by a LD model in parallel
magnetic field $H_{\parallel}$, with a heavily renormalized anisotropy
parameter $\gamma (T)$ that diverges exponentially as  $T$
approaches the decoupling transition temperature $T_*$ from below.
The renormalization down of the inter-layer coupling
is due physically to the excitation of vortex rings$^5$ (fluxons)
that lie in between consecutive layers.

Second, we compute the
line tension of a single Josephson vortex in the 
simplest case of an isolated weakly coupled double-layer $XY$-model,
where we find that the parallel lower-critical
field $H_{c1}^{\parallel} (T)$ of the double-layer vanishes exponentially as
$T$ approaches $T_*$ from below.
This result agrees up to a numerical factor 
with a recent calculation by the author of the same
that employs an alternate ``frozen''
superconductor description of the Meissner phase
in layered superconductors.$^8$
Related results have also been obtained by Browne and Horovitz 
in the setting of long
Josephson junctions$^{18}$ and by Horovitz via a
fermion analogy for layered superconductors in parallel magnetic
field.$^6$
The former coincidence is not surprising since the present double-layer
type-II superconductor is equivalent to a (dynamical) 
long Josephson junction at zero
temperature,$^{19-22}$ wherein
no current is allowed to pass   between the junction.
In fact, the above analysis proceeds by
first considering the length of the vortex as imaginary time. 
Semi-classical
quantum corrections to the energy of the fundamental sine-Gordon
soliton,$^{23-25}$ which corresponds to the Josephson vortex in the double
layer, are then computed.  Entropic wandering    of the vortex 
therefore translates
into quantum fluctuations of the soliton.  Note that the wandering of
Josephson vortices is equivalent to the excitation of double-layer
fluxons, which are again the physical origin of this phenomenon.

We next consider a one-dimensional 
lattice of Josephson vortices in the double
layer.  After  adapting certain elements from the analysis  of
long Josephson junctions in external 
field$^{19-22}$ to
the ``semi-classical'' analysis discussed above in
the case of a periodic array of sine-Gordon solitons, we 
are able to compute the reversible 
magnetization as a function of parallel magnetic field.
Notably, we
obtain a cross-over field $B_*^{\parallel}(T) 
\sim  \Phi_0/d^2 \gamma (T)$,
beyond which the magnetization displays a $B_{\parallel}^{-3}$ tail
characteristic of both long Josephson junctions$^{19}$ and of
layered superconductors in high parallel field,$^{26}$ generally
(see Fig. 1 and ref. 27).  
Unlike long Josephson junctions, however, this cross-over
field is much larger than the lower-critical field.
Thus we find no evidence for field dependence in the
layer decoupling transition temperature, $T_*$, within the present
``semi-classical'' approximation, which is in agreement with the results of
Korshunov and Larkin.$^{14}$  Last, we argue that such is the case
for any  finite number of layers, since the isolated double
layer represents the weakest link.

The remainder of the paper is organized as follows:  in part II
we introduce the frustrated $XY$ model in the Villain form,
from which we derive the renormalized LD model for
$T<T_*$.  The double-layer case is the focal point of
section III, 
where we compute the parallel lower-critical field 
and the parallel reversible magnetization from the above
LD model, in addition to 
the compressibility modulus of the corresponding
vortex array and the effective
inter-vortex interaction potential in the dilute limit.
We then apply these results to the phenomenology of
layered type-II superconductors in section IV,  
as  summarized by Fig. 2.
Last, we assess the validity of the present ``semi-classical''
approximation, as well as discuss the general case of $N$
layers, in section V.

\bigskip
\bigskip
\centerline{\bf II. Frustrated $XY$ Model}
\bigskip

The object now is to understand extreme type-II superconducting
films composed of a finite number $N$  of weakly coupled layers in
the presence of external magnetic field.  In the intermediate regime
of the mixed phase, where the magnetic field satisfies
$H_{c1}\ll H \ll H_{c2}$, the London penetration length is in
general much larger than the inter-vortex spacing.
Following Li and Teitel,$^{10}$  magnetic
screening effects are then negligible, and we may describe the system by
a uniformly frustrated layered $XY$ model with an energy
functional given by
$$\eqalignno{ E_{XY} = J_{\parallel}&
\sum_{l=1}^{N}\sum_{\vec r}\sum_{\mu=x,y}
\{1-{\rm cos}[\Delta_{\mu}\phi(\vec r,l)-A_{\mu}(\vec r, l)]\}\cr
&+J_{\perp}\sum_{l=1}^{N-1}\sum_{\vec r}
\{1-{\rm cos}[\phi(\vec r, l+1)-\phi(\vec r, l)-A_z(\vec r, l)]\}.
& (1) \cr}$$
Here $\phi(\vec r, l)$ is the phase of the superconducting
order-parameter on the layered structure, where
$\vec r$ ranges over the square lattice with
lattice constant $a$,  and    $l$ is the index
for the layers separated by a distance $d$.  
We presume that the lattice constant $a$ is larger than the
size of a typical Cooper pair.
The magnetic flux threading the
plaquette at site $(\vec r, l)$  perpendicular to the
$\mu = x,y,z$ direction reads
$\Phi_{\mu} = (\Phi_0/2\pi) \sum_{\nu,\gamma} \epsilon_{\mu\nu\gamma}
\Delta_{\nu} A_{\gamma}$, where
$\Phi_0 = hc/2e$ is the flux quantum.
Also, $\Delta_{\mu}\phi(r)=\phi(r+\hat\mu)-\phi(r)$
is the lattice difference operator.
The nearest-neighbor
Josephson couplings are related to the respective
masses in Ginzburg-Landau theory  by
$$\eqalignno{J_{\parallel} = &(\hbar^2/2m_{\parallel} a^2)(n_s a^2 d), &(2a)\cr
                 J_{\perp} = &(\hbar^2/2m_{\perp} d^2)(n_s a^2 d), &(2b)\cr}$$
where $n_s$ labels the superfluid density.
Notice that $J_{\parallel}$ is independent of the lattice constant,
$a$, as required by scale invariance in two dimensions.
Hence the anisotropy parameter, 
$\gamma^{\prime} = (J_{\parallel}/J_{\perp})^{1/2}$, 
of the $XY$ model is related to that of the
mass, $\gamma = (m_{\perp}/m_{\parallel})^{1/2}$,
by 
$$\gamma^{\prime} = \gamma {d\over a}. \eqno (3)$$
Throughout, we will presume weak coupling between layers,
$\gamma^{\prime}\gg 1$. 
 
To compute the corresponding partition function
$Z=\int {\cal D} \phi (r) e^{-E_{XY}/k_B T}$, we now make
the usual
(low-temperature) Villain substitution  for the exponential 
factors above;$^{15}$ i.e.,
$e^{-\beta (1 - {\rm cos}\,\theta)}\rightarrow
(2\pi\beta)^{-1/2}\sum_{n=-\infty}^{\infty}
e^{i n\theta} e^{-n^2/2\beta}$.   After   
integrating over the phase-field,$^{7}$ we then obtain the
following dual representation equivalent to
$N$-layered compact quantum electrodynamics (PQED)
in the strong-coupling
limit (modulo a constant):$^{28}$
$$\eqalignno{Z = \sum_{\{n_{\mu}(r)\}}\Pi_{r}
\delta\Biggl[\sum_{\nu}\Delta_{\nu} n_{\nu}|_{r}\Biggr]
{\rm exp}\Biggl[-{1\over {2\beta_{\parallel}}}\sum_{l=1}^{N}
\sum_{\vec r} \vec n^2(\vec r,l)&
-{1\over{2\beta_{\perp}}}\sum_{l=1}^{N-1}
\sum_{\vec r}n_z^2(\vec r,l)\cr
&-i\sum_{r,\nu} n_{\nu}(r)A_{\nu}(r)\Biggr],
& (4)\cr}$$
where $n_{\mu}(r)$ is  an integer link-field
on the layered lattice structure  of
points $r=(\vec r, l)$, with   $\mu=x,y,z$ and
$\vec n=(n_x,n_y)$.  Here, $\beta_{\parallel,
\perp}=J_{\parallel,\perp}/k_B T$.  Notice that 
the conserved integer field $n_{\mu}$ is 
conjugate to the superfluid current
$\Delta_{\mu}\phi - A_{\mu}$ of the $XY$ model (1) in the
continuum limit.  To proceed further,
let us now decompose the parallel field $\vec n$ into 
transverse and longitudinal 
parts $\vec n(\vec r, l) = \vec n\,^{\prime}(\vec r, l)
- \vec n_-(\vec r, l) + \vec n_-(\vec r, l-1)$,
where the transverse and longitudinal fields,
$\vec n^{\prime}$ and $\vec n_-$, respectively
satisfy the constraints
$\vec\nabla\cdot\vec n^{\prime} = 0$ and
$\vec\nabla\cdot\vec n_- = n_z$, with $\vec\nabla = (\Delta_x, \Delta_y)$.
We then take the customary potential
representation $\vec n_{-}=-\vec \nabla\Phi$
for the longitudinal (inter-layer) field,
which yields
$\Phi(\vec r, l) =
\sum_{\vec r\,^{\prime}} G^{(2)}(\vec r - \vec r\,^{\prime})
n_z(\vec r\,^{\prime}, l)$, where 
$$G^{(2)}(\vec r)= 
\int_{\rm BZ}{d^2k\over{(2\pi)^2}} {e^{i\vec k\cdot \vec r}\over
{4-2 {\rm cos} (k_x  a) -2 {\rm cos} (k_y a)}}\eqno (5)$$ 
is (formally) the Greens function for the square lattice.
In the limit
of weak coupling, $\gamma^{\prime}\rightarrow \infty$, the
interlayer field $\vec n_-$ vanishes, which implies that
$\vec n^{\prime}$ is indeed an integer field.
After making a  suitable (lattice)
integration by parts of the energy functional in Eq. (4),
we then obtain the factorization
$Z = Z_{\rm CG}\Pi_{l=1}^{N} Z_{\rm DG}^{(l)}$ for
the partition function in the limit
of weakly coupled layers, where 
$$Z_{\rm DG}^{(l)} = \sum_{\{\vec n^{\prime}(\vec r, l)\}}\Pi_{\vec r}
\delta[\vec\nabla\cdot\vec n^{\prime}|_{\vec r, l}]
{\rm exp}\Biggl[-{1\over {2\beta_{\parallel}}}
\sum_{\vec r} \vec n^{\prime 2}(\vec r,l)
-i\sum_{\vec r} \vec n^{\prime}(\vec r,l)
\cdot\vec A(\vec r, l)\Biggr],\eqno (6)$$
is the partition function for  the 2D discrete gaussian model,$^{29}$
with the in-plane vector potential $\vec A = (A_x, A_y)$
presumed to be  in the gauge $\vec\nabla\cdot\vec A = 0$,
while the inter-layer Coulomb gas factor reads
$$\eqalignno{
Z_{\rm CG} =  \sum_{\{ n_z(\vec r, l)\}} {\rm exp}
\Biggl\{ -{1\over{2\beta_{\parallel}}}\sum_{l=1}^{N}
\sum_{\vec r, \vec r\,^{\prime}}[&n_z(\vec r, l-1) - n_z(\vec r, l)] 
G^{(2)}(\vec r - \vec r \,^{\prime})
[n_z(\vec r\,^{\prime}, l-1) - n_z(\vec r\,^{\prime}, l)]\cr
&-i\sum_{l=1}^{N-1}\sum_{\vec r} n_z(\vec r, l) A_z(\vec r, l)
-{1\over{2\beta_{\perp}}}\sum_{l=1}^{N-1}
\sum_{\vec r} n_z^2(\vec r, l)\Biggr\},& (7) \cr}$$
with the fields at the boundary layers set to
$n_z(\vec r, 0) = 0 = n_z(\vec r, N)$.  
This factorization is consistent with the original layered $XY$ model
(1) that consists of $N$ independent 2D $XY$ models in this limit.
For the more relevant case of $\gamma^{\prime}$ large compared to one,
but finite, we now make note of the fact that these $XY$ layers
remain   effectively 
decoupled in the presence of perpendicular magnetic fields 
that are larger
than the Glazman-Koshelev decoupling scale,$^{16}$
$B_*^{\perp}\sim \Phi_0/\lambda_J^2(0)$, where
$$\lambda_J  (0) = \gamma^{\prime} a = \gamma d \eqno (8)$$ 
is the Josephson penetration length.
We therefore argue on
a physical basis that the above factorization prevails 
in the presence of Josephson coupling as long
as the perpendicular field satisfies $H_{\perp}\gg B_*^{\perp}$
in thin films of layered superconductors.

In the absence of external magnetic field, 
$A_{\mu}(r)=0$, each layer (6) thus
undergoes a
KT transition at $k_B T_c\lsim {\pi\over 2} J_{\parallel}$,
while the inter-layer links $n_z(\vec r)$ undergo an
inverted  2D Coulomb gas
binding transition at $k_B T_*= 4\pi J_{\parallel}$ in the limit
of weak inter-layer coupling, $\gamma^{\prime}\rightarrow\infty$.$^7$ 
(It is understood that the limit of vanishing Josephson
coupling is taken before that of vanishing
perpendicular field.)
The latter high-temperature
transition,  which 
occurs well inside the normal phase  of
each individual layer, corresponds to
the decoupling of layers
mediated by the binding of oppositely ($n_z$)
charged  vortex rings 
lying in between consecutive layers.$^{5}$  Notice 
this implies that
Josephson coupling between {\it resistive} layers exists
in the temperature regime   $T_c < T < T_*$!$^{8,11,12}$
For $\gamma^{\prime}$ large but finite, 
the form (7) of the layered Coulomb gas
ensemble indicates that  each set of
consecutive double layers is 
dielectrically screened by itself as well as by the neighboring
$N-2$  such double layers.  Hence, they each can be
considered in isolation  from their neighbors as long as one
makes the replacement
$\beta_{\parallel} \rightarrow \epsilon_0^{N-1} \beta_{\parallel}$,
where $\epsilon_0 - 1$
is the polarization of an isolated double layer. 
Since the latter is
directly proportional to the fugacity,
$y_* = {\rm exp}(-2\pi\gamma^{\prime 2})$, of the
Coulomb gas (7) at $T_*$ [see Eq.(11)],$^{8}$  and since
the decoupling transition temperature is then given by
$k_B T_*\cong 4\pi J_{\parallel}[1+(N-1)(\epsilon_0-1)]$
in the limit of weak inter-layer coupling, we 
obtain an implicit linear dependence
for the corrections to the value of $T_*$ with the fugacity $y_*$. 
Such a linear dependence agrees with the
standard renormalization-group flows that correspond to the 2D
Coulomb gas.$^{30}$
As expected, the above  formula for $T_*$
also indicates that the decoupling transition
temperature increases without bound with the  number of layers.
In particular,
the  former linear increase crosses over to an exponetial increase
at  $N_0\sim(\epsilon_0-1)^{-1}$ layers, which is far
beyond the 2D-3D cross-over point expected to occur at 
$N_{\rm c/0}\sim \gamma^{\prime}$ 
layers for the layered $XY$ model.$^{31}$
This is then consistent with the fact that the {\it bulk}
layered $XY$ model exhibits only a $3D$ superfluid transition
at the bulk  $T_c$.  Note that the present
factorization into parallel and perpendicular parts is unable to
obtain corrections for the value of $T_c$ in  the case of
large but finite anisotropy parameters
$\gamma^{\prime}$.$^{32}$

Consider now Eqs. (6) and (7) in the presence of 
a homogeneous magnetic induction, 
$$\eqalignno{
B_{\parallel} & = {\Phi_0\over{2\pi  d}} b_{\parallel}, & (9a)\cr
B_{\perp} & = {\Phi_0\over{2\pi a}} b_{\perp}, & (9b)\cr}$$
with the parallel component directed along the $y$-axis,
and with  $B_{\perp}\gg B_*^{\perp}$ to insure the decoupling
between 2D perpendicular  vortices and parallel Josephson
vortices explicit in the previous factorization of (4).$^{16}$
This decoupling becomes evident if we choose   the
gauge $A_x = 0$, $A_y = b_{\perp} x$, and 
$A_z = - b_{\parallel} x$, where 
$b_{\parallel}$ and $b_{\perp}$ are 
the parallel (7) and perpendicular (6) magnetic flux
densities.
In particular, each layer independently experiences the perpendicular 
component $B_{\perp} = H_{\perp}$ of the magnetic
induction,$^{33}$ which sets the intra-layer   vortex density
to be $n_V = |H_{\perp}|/\Phi_0$.  The fact that     
$n_V \gg\lambda_J^{-2}\gg \lambda_L^{-2}$ in the present limit of extreme
type-II superconductivity
insures that magnetic screening
effects transverse to the perpendicular field component can be 
neglected.
Each layer will then independently follow the 2D
vortex lattice melting scenario,$^{17}$ with a melting
temperature $T_m < T_c$.
Since  the issue of 2D vortex lattice melting
has already been discussed
extensively in the literature with
respect to the phenomenon of high-temperature
superconductivity,$^1$ we shall end the present discussion
here and 
focus our attention below on the thermodynamics
connected with the parallel component to the magnetic induction.

We now derive the renormalized LD theory$^{2-4}$
mentioned in the introduction.  It is useful first 
to make the following Hubbard-Stratonovich transformation of
the  Coulomb gas ensemble (7):$^{34}$
$$\eqalignno{
Z_{\rm CG} = \int{\cal D}\theta(\vec r,l) 
\sum_{\{n_z(\vec r, l)\}} {\rm exp}\Biggl\{&
-{\beta_{\parallel}\over 2}\sum_{l = 1}^{N}\sum_{\vec r}(\vec\nabla\theta)^2
- i \sum_{l = 1}^{N}\sum_{\vec r}\theta (\vec r, l)[n_z(\vec r, l) - 
n_z(\vec r, l-1)]\cr
&-i \sum_{l = 1}^{N-1}\sum_{\vec r}n_z (\vec r, l) A_z(\vec r,l)
-{1\over{2\beta_{\perp}}}\sum_{l = 1}^{N-1}\sum_{\vec r} n_z^2(\vec r, l)
\Biggr\}. &(10)\cr}$$
Here $\theta (\vec r, l)$ represents a real scalar field that 
lives on each layer.  
Suppose now that we operate in the low-temperature regime,
$T < T_*$, where the layers are coupled.$^{5-7}$
Then inter-layer $n_z$ charges in the Coulomb gas ensemble are
screened, which means that global charge conservation is no longer
enforced.  Following the standard prescription,$^{34}$
we independently sum over charge configurations at each site,
with the
restriction  to values $n_z = 0, \pm 1$.  In the limit that the fugacity
$$y_0 = {\rm exp}( - 1/2\beta_{\perp})\eqno (11)$$ 
is small,
we then find   that the original Coulomb gas ensemble (7) is 
equivalent to a renormalized Lawrence-Doniach model
$Z_{\rm CG} =\int{\cal D}\theta (\vec r, l) {\rm exp}(-E_{\rm LD}/k_B T)$
with energy functional
$$\eqalignno{
E_{\rm LD} = 
J_{\parallel}\sum_{l = 1}^{N}\sum_{\vec r}
{1\over 2}(\vec\nabla\theta)^2
+2y_0 k_B T\sum_{l = 1}^{N-1}\sum_{\vec r}\{1-{\rm cos}[\theta(\vec r,l+1)-
\theta (\vec r,l)-A_z(\vec r, l)]\}.\cr
&&(12)\cr}$$
At the decoupling transition in particular, 
we have that the fugacity (11) is given by
$y_* ={\rm exp}(-2\pi\gamma^{\prime 2})$.  Hence, the anisotropy parameter
is renormalized up to 
$$\gamma_*^{\prime} = \Biggl({J_{\parallel}\over{2 y_* k_B T_*}}\Biggr)^{1/2} 
= (8\pi)^{-1/2}e^{\pi\gamma^{\prime 2}}\eqno (13)$$
at the decoupling transition
in the present LD model.  Since the $n_z$ charges physically
represent vortex rings  (fluxons)
that lie in between consecutive layers,$^5$
we conclude that such excitations are responsible
for the renormalization (13) of the mass anisotropy
near the decoupling transition.   In closing, we  remind
the reader that the above derivation of model (12) is valid
only for fugacities (11) that satisfy $y_0\ll 1$; i.e., for temperatures 
$T > J_{\perp}/k_B$.

\bigskip
\bigskip
\centerline{\bf III. Double Layer}
\bigskip

We shall now consider the parallel thermodynamics 
associated with the renormalized
LD model (12) in the special case of a double layer ($N=2$),
which is analytically tractable. 
This case is very similar to a long Josephson
junction$^{19-22}$ restricted to
pass no current between the junction.
Although the vortex lattice
that results     from the infinite-layer LD model 
in parallel field
has far more structure than the simple
vortex array corresponding to
an isolated double layer,$^{26}$ we believe that it  is sufficient to
study the latter with respect to the issue
of layer decoupling in general, since it represents the weakest
link.

For the special case of two weakly coupled layers in 
the presence of a homogeneous parallel
magnetic induction $B_{\parallel}$
directed along the $y$-axis, the partition function
corresponding to the renormalized
LD model energy functional (12) is expressible as
$$Z_{\rm CG} = 
\int{\cal  D}\bar\theta(\vec r) {\cal D} \phi(\vec r\, ^{\prime})
{\rm exp}\, \Biggl\{-\bar\hbar_F^{-1}\int dy L_{\rm SG}[\phi]  -
\int d^2r  {\beta_{\parallel}\over 2}
(\vec\nabla\bar\theta)^2\Biggr\},
\eqno (14)$$
where $\bar\theta(\vec r) =  2^{-1/2}[\theta(\vec r, 1)+\theta(\vec r, 2)]$.
Here,
$$L_{\rm SG}[\phi] = \int dx\Biggl[{1\over 2}\Biggl({\partial\phi  \over
{\partial y}}\Biggr)^2 + {1\over 2}\Biggl({\partial\phi  \over  
{\partial x}} - b_{\parallel}\Biggr)^2  + \Lambda_0^{-2}
(1-{\rm cos}\, \phi  )\Biggr]\eqno (15)$$
represents the ``Lagrangian'' for the sine-Gordon model in one space ($x$)
and one imaginary time ($y=i\bar t$) dimension, with a 
bare Josephson penetration length 
$$\Lambda_0 = a (\beta_{\parallel}/4 y_0)^{1/2},\eqno (16)$$
while the effective dimensionless Planck constant is
$$\bar\hbar_F = 2/\beta_{\parallel}.\eqno(17)$$
Notice  that we have taken the continuum limit of the LD
model (12), as well as  made the change of variable
$\phi(\vec r) = \theta(\vec r, 2)-\theta(\vec r, 1) - A_z(\vec r)$.
The integration over $\bar\theta$ on the
right of Eq. (14) results in a trivial 
gaussian factor.  Below, we shall exploit the quantum mechanical
analogy  suggested above for the  nontrivial sine-Gordon factor
in order to compute the parallel lower-critical
field$^{8}$ and the parallel
reversible magnetization of the double-layer system.

\bigskip
\centerline{\bf A. Single Josephson Vortex}
\bigskip
We now set ourselves to the task of
computing  the parallel lower-critical
field, $H_{c1}^{\parallel}$, of the double layer system, which
is in general related to the free energy per unit length of
a single Josephson vortex, $\varepsilon_{\parallel}$, by$^1$
$H_{c1}^{\parallel} = 4\pi\varepsilon_{\parallel}/\Phi_0$.
Let us therefore consider the effective sine-Gordon model (15)
in the presence of a single Josephson vortex aligned
along the $y$-axis; i.e., the homogeneous
magnetic flux is set to 
$b_{\parallel} = 0$, while the phase-difference field
winds once                  along the $x$-axis;
$\int_{-\infty}^{\infty} dx \partial\phi/\partial x = 2\pi$.
In the absence of thermal (or ``quantum'') fluctuations,
the vortex line tension is given by the Ginzburg-Landau energy 
$$\varepsilon_{\parallel}^{0} = {J_{\parallel}\over 2} L_{\rm SG}[\phi_0]
= {4J_{\parallel}\over{\Lambda_0}}\eqno (18)$$
of the ``static'' fundamental sine-Gordon soliton
$$\phi_0(x, y) = 4\, {\rm tan}^{-1} e^{x/\Lambda_0},\eqno (19)$$ 
which is a solution of the field equation
$$-{\partial^2\phi\over{\partial x^2}} + \Lambda_0^{-2} {\rm sin}\, \phi = 0
\eqno (20)$$
obtained by minimizing the corresponding ``action'' 
$L_y L_{\rm SG}[\phi_0]$,$^{23,25}$
and which represents the single Josephson vortex.
For temperatures near the decoupling transition at $T_*$, however,
vortex wandering is critical,$^{8}$ and entropic (or ``quantum mechanical'')
corrections to the vortex line energy (18) must be accounted
for.

We shall now include the effects of thermal wandering in the double-layer
Josephson vortex (19) by first Wick rotating the $y$ coordinate
to a  real time-like coordinate, $y=i\bar t$.  Second, we 
observe that
{\it the free energy per unit length of the Josephson vortex is equal to
the product of ${1\over 2}J_{\parallel}$ with
the ``quantum mechanically'' renormalized ``mass'' of the fundamental
sine-Gordon soliton}.  To obtain the latter, we shall employ the
``semi-classical'' approximation$^{24,25}$ generally valid
in the limit $\bar\hbar_F\rightarrow 0$; i.e., at low temperature.
In particular, consider small deviations
$\phi = \phi_0 + \phi_1$ from the
``static'' vortex configuration (19).  Then 
integration by parts yields that (15) is approximately
$$L_{\rm SG}[\phi] \cong L_{\rm SG}[\phi_0] +
{1\over 2}\int dx \phi_1^*
\Biggl({\partial^2\over{\partial\bar t^2}} -
{\partial^2\over{\partial x^2}} + \Lambda_0^{-2}{\rm cos}\, \phi_0\Biggr)
\phi_1\eqno (21)$$
to second order in the deviation.  Hence in the presence of the soliton, 
we obtain a spectrum of harmonic
oscillators of
the form $\phi_1(x,\bar t) = \psi  (x) e^{i\omega\bar t}$, where
$$\Biggl(-{\partial^2\over{\partial x^2}} + 
\Lambda_0^{-2}{\rm cos}\, \phi_0\Biggr)
\psi   = \omega^2\psi.\eqno (22)$$
It is well known$^{23,25}$ that the spectrum 
corresponding to (22) is composed of
a zero-mode
$$\psi_b(x) = {\rm sech}\, (x/\Lambda_0), \quad\omega_b = 0 \eqno (23)$$
that lies within the gap  of the continuum
$$\psi_k(x) =  e^{ikx}[k + i\Lambda_0^{-1}{\rm tanh}(x/\Lambda_0)],
\quad\omega_k = (k^2 + \Lambda_0^{-2})^{1/2}.\eqno (24)$$
In the ``semi-classical'' approximation,$^{25}$
the ``transition amplitude'' over 
a period of ``time'' $\bar T_0$ is
given by the product $Z_{\rm SG}[1] = {\rm exp}
(-i\bar T_0\varepsilon_{\parallel}^0/k_BT) \Pi_k z_k$, where
$$z_k = \sum_{n=0}^{\infty} {\rm exp}\Biggl[-i\omega_k\bar T_0
\Biggl(n+{1\over 2}\Biggr)\Biggr]
\eqno (25)$$
gives the corresponding amplitude of each harmonic oscillator.
After Wick rotating back to imaginary time $L_y = i\bar T_0$,
we observe that only the $n=0$ terms above survive
the limit of a long vortex, $L_y\rightarrow\infty$.
Yet the ratio of the partition functions in the presence 
of a single Josephson  vortex to that in its absence
is in general related to the vortex line tension,
$\varepsilon_{\parallel}$, by
$Z_{\rm SG}[1] / Z_{\rm SG}[0] =
{\rm exp} (- L_y\varepsilon_{\parallel}/k_BT)$.
In the ``semi-classical'' limit, therefore,
we obtain   
$$\varepsilon_{\parallel} = \varepsilon_{\parallel}^0
+{k_BT\over 2} \Biggl(\sum_{k} \omega_k
- \sum_q \omega_q\Biggr)\eqno (26)$$
for the line tension,$^{24,25}$ 
where the $k$ sum and the $q$ sum above correspond respectively to
traces of the zero-point energy in the presence and in the absence
of the fundamental sine-Gordon soliton.  
In particular, the presumption of periodic boundary conditions along
$x$ implies the quantization conditions
$k L_x + \delta(k) = 2\pi n$ and $q L_x = 2\pi n$, 
where
$$\delta(k) = 2\, {\rm tan}^{-1} (1/k\Lambda_0) \eqno (27)$$
is the phase shift of the continuum states (24),
and where  $n$ is any integer.
Properly counting these 
states then yields that the difference in brackets,
$\sum_n (\omega_k -\omega_q)$, in Eq. (26)
is equal to$^{24,25}$
$-(2\pi)^{-1}\int_{-\infty}^{\infty} dk (d\omega_k/dk)\delta(k)$
in the limit $L_x\rightarrow\infty$.
After introducing
a momentum cut-off, $r_0^{-1}$, 
and integrating by parts, we obtain
$$\varepsilon_{\parallel} = \varepsilon_{\parallel}^0
[1 - (4\pi\beta_{\parallel})^{-1} 
{\rm ln}(\Lambda_0/r_0)].\eqno (28)$$
Notice that the correction above to the low-temperature
line tension is first order in $\bar\hbar_F$, and that it
can be interpreted as a renormalization to the ``mass'',
$\Lambda_0^{-1}$, of the sine-Gordon model
(15).  Since a renormalization group
exists,  we may now express the
line tension as
$$\varepsilon_{\parallel} = 4J_{\parallel}/\lambda_J,\eqno (29)$$
and then iterate      Eq. (28), which  yields
$\lambda_J^{-1} = 
\Lambda_0^{-1}{\rm exp}[-(4\pi\beta_{\parallel}\epsilon_0)^{-1}
{\rm ln}(\lambda_J/r_0)]$,  or
$${r_0\over{\lambda_J}} = \Biggl({r_0\over{\Lambda_0}}\Biggr)^
{[1-(4\pi\beta_{\parallel}\epsilon_0)^{-1}]^{-1}}.\eqno (30)$$
Above, we have replaced the right-hand side of Eq. (28) by the
previous exponential and included the  dielectric correction
$\epsilon_0$ to
the  on-site Coulomb-gas (7) potential, ${\rm ln}(\lambda_J/r_0)$.
Employing the  standard renormalization group result,$^{30}$ 
$\epsilon_0 = 1+{\rm O}[(\beta_{\parallel}-\beta_*)^{1/2}]$,
for the dielectric constant of the
2D Coulomb gas at
(inverted) temperatures $\beta_{\parallel}$ just above 
the (inverted) transition temperature
$\beta_*=1/4\pi$,
we thus obtain that the renormalized Josephson penetration length 
$\lambda_J$
diverges exponentially as it approaches the
decoupling transition like
$$\lambda_J/a = 
C\, {\rm exp}[D/(\beta_{\parallel}-\beta_*)^{1/2}].\eqno (31)$$
Here, $C$ and $D$ are non-universal numerical constants.

In conclusion, the parallel lower-critical field 
$H_{c1}^{\parallel} = 4\pi\varepsilon_{\parallel}/\Phi_0$
vanishes exponentially fast
near the decoupling transition of the double layer 
following Eqs. (29) and (31).  
Horovitz has obtained this result employing
a fermion analogy for layered superconductors in
parallel magnetic field.$^6$
A similar dependence has also
been proposed by Browne and Horovitz for the
lower-critical field of long Josephson junctions.$^{18}$
Combining  Eq. (2a) with the identity
$(\Phi_0/\lambda_L)^{2} = 
(2\pi)^3(\hbar^2/2m_{\parallel}) n_s$ 
for the bulk ($N\rightarrow\infty$) in-plane
London penetration length $\lambda_L$ yields
$${J_{\parallel}\over d} = (2\pi)^{-3}{\Phi_0^2\over{\lambda_L^2}}, \eqno (32)$$
from which we obtain the useful expression
$$\varepsilon_{\parallel}(T) = {8\over\pi}\gamma^{-1}
\Biggl[{\Phi_0\over{4\pi\lambda_L(T)}}\Biggr]^2
{\lambda_J (0)\over{\lambda_J (T)}}.\eqno (33)$$
Given that $C\sim\gamma^{\prime}$ in Eq. (31),  which is consistent
with expression  (8) for the Josephson penetration length of the original
$XY$ model (1) at zero temperature, then 
we have  by Eq. (31) that
$\lambda_J (0)/\lambda_J (T) \sim {\rm exp} [-D/(\beta_{\parallel} -
\beta_*)^{1/2}]$ near the decoupling transition.
This  result
agrees up to a numerical constant
with a previous calculation 
by the author of the same quantity using an alternative ``frozen''
superconductor model for highly anisotropic
extreme type-II superconductors in the Meissner phase.$^8$
Notice then that expression (33) for the parallel line tension
is essentially independent of the lattice constant $a$,
as expected by the 2D scale invariance of the LD model (1)
for $T\geq T_*$. 

Before we go on to consider an array of Josephson vortices
in the next section, a few remarks are called for.
First, notice that the point at which the soliton ``mass'',
$2\varepsilon_{\parallel}/J_{\parallel}$, vanishes coincides with
the layer decoupling transition.  Similar effects are
found in the case of the nonlinear $\sigma$ model in two space
and one time dimensions, which describes the quantum 2D
antiferromagnet. In particular, the quantum 
mechanically renormalized energy of the corresponding
topological soliton called a skyrmion
vanishes precisely at the zero-temperature
quantum critical phase transition into the
quantum disordered phase characterized by a spin-gap.$^{35}$
Second, notice that we have essentially recovered the
standard renormalization group results for the KT
transition$^{30}$ via the present semi-classical
quantization of the sine-Gordon soliton energy in 
one space and one time dimension.  Finally, 
also observe that the entropic
correction to the line-tension in Eq. (28) indicates that the number
of microstates per unit length $a$ of a Josephson vortex 
in thermal equilibrium is
$\lambda_J/r_0$.  Given that $r_0\sim a$, then $\lambda_J$ 
can be naturally 
interpreted as the effective width of the Josephson vortex.

\bigskip
\centerline{\bf B. Array of Josephson Vortices}
\bigskip

Consider now the case of a  nonzero homogeneous magnetic
induction aligned parallel 
to the $y$-axis of the double layer; i.e., $b_{\parallel}\neq 0$.
Then it is easily seen from Eq. (15)
that the superfluid portion,
$$G_s-G_n = -k_B T\, {\rm ln}\int{\cal D}\phi 
e^{-\bar\hbar_F^{-1}\int dy L_{\rm SG}[\phi]},\eqno (34)$$ 
of the Gibbs free energy$^{36}$ is minimized with respect to $b_{\parallel}$
at $b_{\parallel} = L_x^{-1}\int_{-\infty}^{\infty}\partial\phi/\partial x$.
In other words, the average  winding per unit length in any configuration
of  the phase difference between the double-layers is
set by the magnetic induction.  In the particular case
of the low-temperature ``classical'' configuration, we then have  
that the parallel magnetic induction is related to 
the lattice constant $a_0$  of the corresponding
array of Josephson vortices by
$$b_{\parallel} = {2\pi\over{a_0}}.\eqno (35)$$
Clearly, we expect qualitative differences 
between the thermodynamics of the
the low-field regime, $a_0\gg\lambda_J$, and of the high-field regime,
$a_0\ll\lambda_J$.  Yet 
does a decoupling phase transition separate the two
regions, as has been suggested in the literature?$^{3}$
Below we give evidence for the existence of only a crossover$^{14}$ 
on the basis of a ``semi-classical''
analysis of the reversible magnetization
(see ref. 27) and of the elastic
compression modulus of the vortex array.

We  now set ourselves to the task of computing 
the parallel reversible magnetization$^{36}$
$$M_{\parallel} \cong  - {\partial\over{\partial B_{\parallel}}}
\Biggl({G_s-G_n\over{L_x L_y d}}\Biggr)\eqno (36)$$ 
of the double  layer in the intermediate regime of the
mixed phase,
$H_{c1}^{\parallel}\ll B_{\parallel}\ll H_{c2}^{\parallel}$, where
$H_{\parallel}\cong B_{\parallel}$.
Consider first the lowest order Ginzburg-Landau contribution
$$G_s^0-G_n = L_y {J_{\parallel}\over 2} L_{\rm SG}[\phi_0]\eqno (37)$$
to the Gibbs free energy in powers of $\bar\hbar_F$,
where $\phi_0(x)$ represents the whirling pendulum solution of 
field Eq.
(20) with spatial period $a_0$; i.e.,$^{19-22}$
$${d\phi_0\over{dx}} = {2\over{\kappa\Lambda_0}}
{\rm dn}\Biggl({x-x_0\over{\kappa\Lambda_0}}\Biggl|\kappa^2\Biggr),\eqno (38)$$
where the parameter $\kappa$ lies in the interval
between zero and unity, and is  set by the period $a_0$ following
$$a_0 = 2\Lambda_0\kappa K(\kappa^2).\eqno (39)$$
Above, ${\rm dn}(u|\kappa^2)$ represents the appropriate
Jacobian Elliptic function,
while $K(\kappa^2)$ represents the complete Elliptic integral of
the first kind.$^{37}$  
Conservation of energy,
$E_0= 2\Lambda_0^{-2}(\kappa^{-2}-1)$,
in the corresponding pendulum system  yields 
$${L_{\rm SG}[\phi_0]\over{L_x}} 
= 2\Biggl[a_0^{-1}\int_0^{a_0} dx {1\over 2} 
\Biggl({d\phi_0\over{d x}}\Biggr)^2\Biggr]
-E_0-{1\over 2}b_{\parallel}^2.$$
We obtain, therefore, that the zero-order Gibbs free energy
density is equal to
$${G_s^0-G_n\over{L_x L_y}} = 
{J_{\parallel}\over{\Lambda_0^{2}}}\Biggl[{2\over{\kappa^{2}}}
{E(\kappa^2)\over{K(\kappa^2)}} +1-\kappa^{-2}\Biggr]
-{J_{\parallel}\over 2}{b_{\parallel}^2\over 2},\eqno(40)$$
where $E(\kappa^2)$ represents the complete Elliptic integral
of the second kind.$^{37}$  Standard manipulations$^{19}$
then yield that the reversible magnetization (36) is given by
$$-4\pi M_{\parallel} = H_{c1}^{\parallel}
\Biggl[{E(\kappa^2)\over{\kappa}}-{\pi^2\over 4}{1\over{\kappa K(\kappa^2)}}
\Biggr],\eqno (41)$$
where $H_{c1}^{\parallel} = 4\pi\varepsilon_{\parallel}^0/\Phi_0$
is the parallel lower-critical in the Ginzburg-Landau
theory approximation.
Notice that $H_{c1}^{\parallel}$ naturally  sets the maximum value 
of the diamagnetic magnetization (41) at
zero magnetic induction ($\kappa = 1$).  
In particular, at low magnetic inductions
$a_0\gg\Lambda_0$, we have by (39) that $\kappa^2\cong 1-
16 e^{-{a_0/\Lambda_0}}$.
We then  obtain the limiting behavior
$$-4\pi M_{\parallel} \cong H_{c1}^{\parallel}\Biggl[
1+4e^{-{a_0\over{\Lambda_0}}}\Biggl({a_0\over{\Lambda_0}} +1\Biggr)
-{\pi^2\over 2}{\Lambda_0\over{a_0}}\Biggr]\eqno (42)$$
for the reversible magnetization (41).
Hence, the low-field magnetization extrapolates to zero
at 
$$B_{0}^{\parallel} = {2\over{\pi^2}} {\Phi_0\over{\Lambda_0 d}},\eqno (43)$$
which defines a bare  crossover field.  At high-fields
$a_0\ll \Lambda_0$, 
on the other hand, (39) dictates that
$\kappa \cong \pi^{-1}a_0/\Lambda_0$.
We then obtain that
the limiting behavior
for the reversible magnetization (41) is given by
$$-4\pi M_{\parallel} \cong {H_{c1}^{\parallel}\over{64\pi^2}}
\Biggl({a_0\over{\Lambda_0}}\Biggr)^3\eqno (44)$$
in such case.
This implies a $B_{\parallel}^{-3}$ tail at high fields 
$B_{\parallel}\gg B_{0}^{\parallel}$
that is characteristic of long Josephson junctions and
of layered superconductors in general.$^{19,26}$  A plot of result
(41) spanning both the high-field and low-field
limits is shown in Fig. 1.  Notice that the 
bare crossover field
$B_{0}^{\parallel}$ is much larger than the 
Ginzburg-Landau lower-critical field
$H_{c1}^{\parallel}$ in the present double-layer 
extreme type-II superconductor [see Eq. (51)].
This is qualitatively different from the case 
of a long Josephson
junction,$^{19}$ where $B_{0}^{\parallel}\sim H_{c1}^{\parallel}$.

In analogy with the previous analysis of a single Josephson vortex,
let us now consider the effect of ``semi-classical'' corrections
to the reversible magnetization in parallel field (41).
We again have a spectrum (21) of harmonic oscillators 
$\phi_1(x,\bar t) = \psi  (x) e^{i\omega\bar t}$ that satisfy
the linearized field equation (22), but with a periodic
configuration for the phase difference set by$^{20,21}$
$${\rm cos}\,{1\over 2}\phi_0 = 
- {\rm sn}\Biggl({x-x_0\over{\kappa\Lambda_0}}\Biggl|\kappa^2\Biggr),
\eqno (45)$$
where ${\rm sn}(u|\kappa^2)$ represents the appropriate Jacobian
Elliptic function.$^{37}$  To be more specific, the spatial factors
of each oscillator satisfy Lam\' e's equation,$^{21}$
$$\Biggl[-{\partial^2\over{\partial x^2}}
+2\Lambda_0^{-2} 
{\rm sn}^2\Biggl({x-x_0\over{\kappa\Lambda_0}}\Biggl|\kappa^2\Biggr)
-\Lambda_0^{-2}\Biggr]
\psi = \omega^2 \psi.\eqno(46)$$
To make contact with the previous discussion of a single Josephson
vortex, let us now focus our attention
on the (bare) low-field
regime $B_{\parallel}\ll B_{0}^{\parallel}$, where the parameter 
$\kappa$ is exponentially close to unity, since
$a_0\gg\Lambda_0$.  This allows us to 
approximate the potential
terms in Lam\' e's equation 
$[-{\partial^2\over{\partial x^2}} + V(x)]\psi = (\omega^2 - 
\Lambda_0^{-2})\psi$
by
$$V(x)
\cong -\sum_{n = -\infty}^{\infty}
2\Lambda_0^{-2}
{\rm sech}^2\Biggl({x-x_0-na_0\over{\Lambda_0}}\Biggr),
\eqno (47)$$
where each term above corresponds to the potential associated with
a fundamental sine-Gordon soliton centered at $x_0+na_0$.  In general,
the band structure corresponding to Lam\' e's equation (46) is 
composed of a
continuum and a zero-mode band separated by a gap.$^{21,37}$  
A  curious feature particular to each potential term in Eq. (47), however,
is its {\it transparency};$^{23}$ i.e., the continuum oscillators 
(24) of the fundamental sine-Gordon soliton
have no reflected wave component.  Therefore
in the present (bare) low-field limit, the
upper continuum band is         
essentially the same as that of a  fundamental soliton (24).  
Repeating the renormalization group arguments made in the previous section
for the line energy of a single Josephson vortex 
then indicates that the entropic correction due to
the latter continuum band can be accounted for 
by simply replacing $\Lambda_0$ (16)  with $\lambda_J$ (31)
in the original Ginzburg-Landau free energy
of the vortex lattice; i.e.,
$$G_s-G_n = {J_{\parallel}\over 2}
\int dy\int dx\Biggl[{1\over 2}\Biggl({\partial\phi_0  \over
{\partial y}}\Biggr)^2 + {1\over 2}\Biggl({\partial\phi_0  \over
{\partial x}} \Biggr)^2  - {1\over 2}b_{\parallel}^2+ \lambda_J^{-2}
(1-{\rm cos}\, \phi_0  )\Biggr],\eqno (48)$$
with the lattice constant of the vortex array (38)
set by Eq. (35).  In general, however,
the effects of the zero-mode band must also be included in the present
semi-classical analysis.  The corresponding states are given by
the tight-binding anzats $|k_0\rangle = \sum_n e^{i k_0  a_0 n}|n\rangle$
in the present (bare) low-field limit,
where $\langle x|n\rangle = \psi_b(x-x_0-na_0)$ 
is the (normalized)  bound state (23)  of the fundamental
soliton located at the $n^{\rm th}$ well.  The hopping matrix element is
therefore $-t_0 = \langle n|-{\partial^2\over{\partial x^2}} + V(x)|n+1\rangle
=(\omega^2 - \Lambda_0^{-2})\langle n|n+1\rangle$.  But
$\langle n|n+1\rangle\cong 2\pi e^{-a_0/\Lambda_0}$
is much less than unity, which  yields
$t_0\cong (2\pi/\Lambda_0^2)e^{-a_0/\Lambda_0}$.  This means that the
zero-mode band has a spectrum $\omega_0(k_0) = 2 t_0^{1/2}|{\rm sin}
({1\over 2} k_0 a_0)|$ that is exponentially narrow.   
By (25), the zero-mode band results in
an entropic {\it pressure} contribution to the Gibbs free energy
density given by
$$P_0 = {k_B T\over d} L_x^{-1}\sum_{k_0} {1\over 2} \omega_0(k_0)
= {2\over\pi} {k_B T \over{a_0 d}}t_0^{1/2}.\eqno (49)$$
Hence, the magnetization (36) acquires a {\it diamagnetic} 
correction $-\partial P_0/\partial B_{\parallel}$ of order
$e^{-a_0/2\Lambda_0}$, which is negligibly small in the present
(bare) low-field limit.  Eq. (42) indicates, however, that the low-field
correction to the initial linear increase of the parallel magnetization
varies as $e^{-a_0/\Lambda_0}$ in the
Ginzburg-Landau regime.  Unlike the case
of a single Josephson vortex, then no obvious renormalization group
appears to exist for the above entropic pressure contribution.

In conclusion,  double-layer extreme type-II superconductors
in parallel magnetic field are    described
by the effective Ginzburg-Landau 
free energy (48),
along with the boundary condition (35), in the
bare low field limit $B_{\parallel}\ll B_{0}^{\parallel}$
of the intermediate regime, $H_{c1}^{\parallel}\ll B_{\parallel}
\ll H_{c2}^{\parallel}$.  This means that
the reversible magnetization is determined 
by the original Ginzburg-Landau theory
analysis [Eqs. (36) - (44)], where    the bare Josephson
penetration length $\Lambda_0$ is replaced by the renormalized length
$\lambda_J$ throughout.  In particular, the true
parallel cross-over field (see Fig. 1) of the double
layer is given by
$$B_*^{\parallel} = {2\over{\pi^2}} 
{\Phi_0\over{\lambda_J d}}\eqno (50)$$
instead of by Eq. (43).  
However, our inability to find a renormalization group for
the entropic pressure contribution (49) to the
parallel magnetization  suggests that the present
renormalized Ginzburg-Landau theory result for $-4\pi M_{\parallel}$
serves only as a strong lower bound in the critical regime.$^{27}$
We therefore find evidence for at best a crossover 
as a function of magnetic field below the  bare scale $B_0^{\parallel}$, 
and no evidence for a decoupling phase transition at
fixed temperature.
Last, it is easily shown after employing relation (32) that
$${B_*^{\parallel}\over{H_{c1}^{\parallel}}} 
= {\lambda_L^2\over{d^2}}.\eqno (51)$$
This  of course indicates that the crossover field is much larger
than the lower-critical field, which 
validates {\it a posteriori} the  assumption (36)
that $H_{\parallel}\cong B_{\parallel}$ in the intermediate
regime of the mixed phase. 
It also illustrates the qualitative difference between a
double-layer superconductor and a long Josephson junction,$^{19}$
where $B_*^{\parallel} \sim H_{c1}^{\parallel}$. 

We shall close this section by computing
the compression modulus of the parallel array of Josephson vortices,
as well as the interaction energy between widely spaced vortices.  
The  local change in the  elastic free-energy density 
due to a local fluctuation $\delta n_V$ in the vortex density is given by 
$\delta f_{\rm SG} = {1\over 2}(\partial^2 f_{\rm SG}/\partial n_V^2)
(\delta n_V)^2$,
where 
$$f_{\rm SG} = 
{J_{\parallel}\over{\lambda_J^{2} d}}\Biggl[{2\over{\kappa^{2}}}
{E(\kappa^2)\over{K(\kappa^2)}} +1-\kappa^{-2}\Biggr]\eqno (52)$$
is the Gibbs free-energy density (48)
modulo the $-{1\over 2}b_{\parallel}^2$ term [see Eq.
(40)], which can be considered
as part of the magnetic field energy.
In general, the number  of Josephson vortices per unit length 
along the $x$-axis is $n_V = a_0^{-1} = b_{\parallel}/2\pi$.
By differentiating the first term in Eq. (41) once
with respect to $b_{\parallel}$, we thus obtain
$\partial^2 f_{\rm SG}/\partial n_V^2 = (8J_{\parallel}/d)
(1-\kappa^2)[K(\kappa^2)]^3/E(\kappa^2)$.  
Now the displacement field $u(x)$ of the vortex array 
along the $x$-axis is related 
to the density fluctuation by
$\delta n_V = - a_0^{-1}(\partial u/\partial x)$.
Therefore employing
previous identities [(9a), (32) and (35)], we find that the elastic energy 
is given by
$\delta f_{\rm SG} = {1\over 2}c_{11}(\partial u/\partial x)^2$,
where the compression modulus reads
$$c_{11} = \pi^{-3}{d^2 B_{\parallel}^2\over{\lambda_L^2}}
{[K(\kappa^2)]^3\over{E(\kappa^2)}}(1-\kappa^2).\eqno(53)$$
In the low-field limit $B_{\parallel}\ll B_*^{\parallel}$
we then have that the array is exponentially
soft,$^{22}$ with $c_{11}\propto e^{-a_0/\lambda_J}$.
Similar softening of the Abrikosov vortex lattice
occurs in conventional superconductors
near the lower critical field,
but with the Josephson penetration length 
$\lambda_J$ replaced by the London penetration
length $\lambda_L$.$^{38}$
On the other hand, 
the high-field limit $B_{\parallel}\gg B_*^{\parallel}$
yields 
that
$c_{11} \cong (4\pi)^{-1}(d/\lambda_L)^2 B_{\parallel}^2$,
which is a formula characteristic  of
the elastic moduli in  extreme
type-II superconductors ($\lambda_L\rightarrow\infty$)
generally.$^1$

Last, we may define the interaction energy between two 
well separated vortices by subtracting the line tension (29)
from the free energy (52) per unit length of a single period
in the array of Josephson vortices; i.e., the 
repulsive interaction
energy per unit length is given by
$$v(a_0) = d a_0 f_{\rm SG} - \varepsilon_{\parallel}\eqno(54)$$
in the low-field limit $a_0\gg\lambda_J$,
which after some analysis yields
$v(a_0) = J_{\parallel}\lambda_J^{-1}(1-\kappa^2)$, or
$$v(x) = 
16 J_{\parallel}\lambda_J^{-1} e^{-|x|/\lambda_J}.
\eqno (55)$$
Notice that the Josephson penetration length $\lambda_J$ 
acts as the screening length for
the interaction between Josephson vortices instead of the
London penetration length, which  plays the same
role in conventional type-II superconductors.

\bigskip
\bigskip
\centerline{\bf IV. Phenomenology}
\bigskip

We shall now examine the phenomenological consequences of
the previous results for (double) layered superconductivity in
parallel external magnetic field.  Let us first consider the
critical properties of the decoupling transition in the
the absence of parallel magnetic induction; i.e., take 
$H_{\parallel}$ near $H_{c1}^{\parallel}$.  Given
the standard Ginzburg-Landau dependence,
$\lambda_L(T) = \lambda_0(1-T/T_{c0})^{-1/2}$, for the 
bulk ($N\rightarrow\infty$) in-plane London
penetration length, then (32) yields
$$J_{\parallel}(T) = \gamma^{-1}
k_B T_0 (1-T/T_{c0})\eqno (56)$$ 
for the intra-layer Josephson
coupling energy, where
$$k_B T_0 = {2\over{\pi}} \Biggl({\Phi_0\over{4\pi\lambda_0}}\Biggr)^2 
\gamma d
\eqno (57)$$
is the  basic energy scale of the problem.  Since the 
zero-field decoupling transition occurs at $k_B T_* = 4\pi J_{\parallel}
(T_*)$, we then obtain
$T_* = [T_{c0}^{-1} + \gamma (4\pi T_0)^{-1}]^{-1}$ for the decoupling
transition temperature.  This means that the size of the critical
regime is 
$$\delta T_* = T_{c0} - T_*\cong \gamma T_{c0}^2/4\pi T_0\eqno (58)$$
for $T_{c0}\ll T_0$, which is typical.
Likewise, the critical temperature $T_c$ at which each individual
layer undergoes a superfluid KT transition is set by
$k_B T_c \cong {\pi\over 2} J_{\parallel}(T_c)$, or
$T_c\cong[T_{c0}^{-1} + \gamma ({\pi\over 2} T_0)^{-1}]^{-1}$.
Hence, the distance of this intra-layer resistive transition
to the Ginzburg-Landau transition temperature $T_{c0}$ is
$\delta T_c = T_{c0} - T_c \sim 10\, \delta T_*$, as indicated
by Fig. 2.  Again, we highlight the extraordinary
regime in temperature $T_c < T < T_*$ where the layers
are normal yet Josephson coupled; i.e., $\rho_{\parallel}\neq 0$
and $H_{c1}^{\parallel}(T)\neq 0$.
This effect has been recently observed in  resistance measurements
on the  highly anisotropic
Bismuth-based 
series of high-temperature superconductors.$^{11,12}$

Results similar to those outlined
above have been obtained recently by the author using
an alternative  anisotropic ``frozen'' superconductor model  for
the Meissner phase,$^8$ but
with the important exception that the
zero-temperature Josephson penetration length  (8)
appearing above in Eq. (57)
is replaced therein by
the zero-temperature London penetration length $\lambda_0$.
This discrepancy can be understood as follows: in the
present frustrated $XY$ model description (1) of
the mixed phase, we take first the limit
$\lambda_0\rightarrow\infty$, and then   the limit
$\gamma\rightarrow\infty$, which  results in the 
characteristic 
Josephson penetration length $\gamma d$.  In the anisotropic ``frozen''
superconductor
model for the Meissner phase,$^8$ on the other hand,
the order of the limits is reversed, hence the
characteristic length scale $\lambda_0$.  
Both models, however, obtain the same expression for the 
parallel lower critical field (33) near
criticality up to a numerical constant.
In particular, we predict that $H_{c1}^{\parallel}(T)$
vanishes exponentially as temperature $T$
approaches the decoupling
transition from below, which implies the
existence of an inflection point 
below $T_*$ in this temperature profile (see Fig. 2). 

In the presence of external parallel magnetic field, we
expect that (double) layered extreme type-II superconductors
follow a cross-over phenomenon as a function 
of this field in the intermediate
regime of the mixed phase.
In particular, Eqs. (39) and (41)
indicate (with $\Lambda_0$ replaced by
$\lambda_J$) that the parallel magnetization has the
form $-4\pi M_{\parallel} = 
H_{c1}^{\parallel}(T) f[B_{\parallel}/B_*^{\parallel}(T)]$,
where the latter functional dependence with parallel magnetic
induction is plotted in Fig. 1.
Notice that in the mixed phase at low magnetic induction, 
$B_{\parallel}\ll B_*^{\parallel}$,
we have that $-4\pi M_{\parallel} \lsim H_{c1}(T)$.
Hence, the parallel magnetization inherits
the inflection of the parallel lower-critical field
as a function of temperature $T\lsim T_*$ at fixed
$B_{\parallel}$.
Fig. 2 shows the phase diagram expected of a (double) layered
superconductor in parallel external magnetic field near the
critical regime discussed above.  
Formula (50) for the crossover field has been employed
here, where
the Josephson penetration length $\lambda_J (T)$ 
is interpolated between its behavior at criticality (31)
and its low	temperature
value of  $\gamma d$.  
Notice that $B_*^{\parallel}$ is expected to be
practically constant 
at low temperatures since the mass anisotropy 
parameter $\gamma$
has no temperature dependence  in this
regime.  We therefore
expect the crossover field to exhibit an inflection point
in its temperature profile, much  like the parallel
lower-critical field does.  
By Eq. (51),  however, the ratio of $B_*^{\parallel}$
to $H_{c1}^{\parallel}$ should be larger at
criticality with respect to zero-temperature
by a factor of $\lambda_L^2(T_*)/\lambda_L^2(0)$,
which in the Ginzburg-Landau theory approximation is
given by $4\pi T_0/\gamma T_* \sim T_{c0}/\delta T_*$.
Last, in spite of the above crossover phenomenon,
the parallel vortex lattice (and flux quantization)
will persist up to the parallel upper-critical
field. In the Landau-Ginzburg approximation, this
field is set by the in-plane coherence length
$\xi = \xi_0(1-T/T_{c0})^{-1/2}$
and by the mass anisotropy to be
$H_{c2}^{\parallel} = \gamma \Phi_0/2\pi\xi^2$,
hence the inequality $B_*^{\parallel}\ll H_{c2}^{\parallel}$.
The critical behavior of $H_{c2}^{\parallel}(T)$
near the decoupling transition at $T_*$, however,  
remains unknown.

We shall now examine the various physical
scales that arise  from
the present theory in the context of high-temperature
superconductivity.  The oxide superconductor
Bi$_2$Sr$_2$CaCu$_2$O$_8$ may be classified
as a layered superconductor with an extreme 
mass anisotropy,$^1$
$\gamma\sim 100$.  
Assuming typical parameters $T_{c0}\sim 100\,{\rm K}$,
$d = 15\,{\rm \AA}$
and $\lambda_0\sim 10^3\,{\rm \AA}$, we obtain from Eq. (58)
that $\delta T_*\sim 0.5\,{\rm K}$.
It is interesting to remark that the zero-temperature
Josephson penetration length $\gamma d$
and the zero-temperature London penetration length, $\lambda_0$,
are of the same order of  magnitude in this material.  The fact that the
present estimate for the size of the critical region is smaller by
an order magnitude with respect to the estimate based on 
the above mentioned anisotropic ``frozen''
superconductor model$^8$ is simply then a result of the 
numerical factor in Eq. (58).  Both theories,
on the other hand, predict an
inflection point in the temperature profile
$H_{c1}^{\parallel}(T)$ (see Fig. 2).
Wan et al. also observe an inflection point in  the
field of first penetration $H_p$ vs. $T$ 
for Bi$_2$Sr$_2$CaCu$_2$O$_8$, 
but in the perpendicular orientation.$^{12}$
This  may be a vestige of the same prediction made here for
the parallel lower-critical field if
geometrical  demagnetization effects
are presumed to be strong.  In particular, 
consider the regime in temperature 
$T_c < T < T_*$, where the planes are resistive while
remaining
Josephson coupled, 
and hence where only parallel Josephson vortices exist.$^8$
Then   the field of first penetration in
the perpendicular orientation
is limited only by those portions
of the field lines that run {\it parallel} to the top and bottom layers  
of the sample.  Clearly, 
direct measurements of the parallel lower
critical field in the critical
regime of these materials would be highly desirable.
Last, we mention that the parallel crossover field (50) at zero temperature,
$B_*^{\parallel}(0) = (2/\pi^2) (\Phi_0/d^2\gamma)$, 
should be approximately
$2$ T for Bi$_2$Sr$_2$CaCu$_2$O$_8$, while it 
should be orders of magnitude
smaller at temperatures
just below the decoupling transition temperature.  We therefore
suggest that the parallel reversible magnetization be measured 
in a clean thin film of this material 
near criticality,$^{27}$ where the crossover field is
expected to be quite modest
in magnitude.  Note that the present theory is valid
only for perpendicular components of magnetic induction
with magnitude greater than the perpendicular crossover
field$^{16}$ $B_*^{\perp}\sim \Phi_0/d^2\gamma^2$, which in the case 
of Bi$_2$Sr$_2$CaCu$_2$O$_8$ is
approximately 1 kG.

\bigskip
\bigskip
\centerline{\bf V. Discussion}
\bigskip

In summary, we find no evidence for field dependence in the 
decoupling  transition
temperature, $T_*$, of clean double-layered 
extreme type-II superconductors in the
intermediate regime of the mixed phase.  Since the double layer represents
the weakest link, we believe that this result remains true in
the general case of any finite number of layers as long as the 
interlayer coupling is weak enough so that the system
remains effectively 2D; e.g., for perpendicular fields
above the Glazman-Koshelev 2D-3D cross-over scale,
$B_{\perp}\sim\Phi_0/\gamma^2 d^2$,
which guarantees the absence of vortex loops 
that traverse many layers.$^{16}$
In general, the extreme type-II
limit $Nd\ll\lambda_L$ must be taken first, however,
so that    magnetic screening effects may be  neglected.  
Also, the effect of pinning centers is  not expected to be relevant
in the critical regime, 
since the Josephson penetration length $\lambda_J (T)$
diverges exponentially as temperature $T$ approaches $T_*$
from below.
We do however find a cross-over
field above which the parallel reversible magnetization decays
with parallel magnetic field $H_{\parallel}$ as $H_{\parallel}^{-3}$ 
in the double-layer case
(see Fig. 1 , Fig. 2, and  ref. 27).  The array of
Josephson vortices is nevertheless 
expected to persist up to the parallel upper-critical field.  
It is important to mention that the present double-layer study
cannot account for effects due to the first order
commensuration transitions predicted to occur in the parallel vortex lattice
with many layers.$^{26}$  

In order to account for  the entropy due to wandering of the
Josephson vortices in the calculations reported
above,   
we have  considered the length dimension of the
vortex as imaginary time, and proceeded to compute the corresponding
``quantum mechanical'' correction to the ``mass''.  Employing
a renormalization-group
improved semi-classical approximation to this end,
we have found that the parallel lower-critical field vanishes exponentially
as it approaches the decoupling transition 
temperature from below.  Very
similar results have been obtained recently by  the author using
an alternative anisotropic ``frozen'' superconductor model
that operates from the Meissner phase.$^8$  Note that although the
dimensionless Planck constant (17) has a value of
$\bar\hbar_F = 8\pi$ at the decoupling
transition, which   is far from being small, it is suspected that
the above cited renormalization-group
improved  semi-classical results are in fact exact
for the case a single Josephson vortex.$^{24,25}$
Less is known, however, with respect to the validity of
the semi-classical approximation in the
critical regime for the case of the array.
For example, we have computed the entropic pressure (49)
of the array to first order in powers of the effective dimensionless
Planck constant (17) and found it to be negligibly small at fields
below a relatively large bare scale (43).  However,
we were unable to find a renormalization group for this
contribution.  Also, it has been argued
in the literature
that the pressure exerted between interfaces in two dimensions
generally varies quadratically with
temperature,$^{39}$ which translates into a 
second order correction in the present semi-classical
approximation.  These effects generally  stiffen the
array of Josephson vortices and add a diamagnetic contribution
to the high-field tail shown by the parallel magnetization (see Fig. 1
and ref. 27).
However, any such additional diamagnetic correction 
does  not affect the conclusion drawn here that no layer decoupling
transition occurs as a function of external magnetic field
in extreme type-II layered superconductors since it 
only results in a {\it stiffer}  vortex lattice.

\bigskip\bigskip

The author is grateful to D. Dominguez, L. Bulaevskii, 
B. Ivlev, M. Maley,
and H. Safar for very useful discussions.  This work
was performed under the auspices of the U.S. Department of Energy,
and  was supported in part by
National
Science Foundation grant DMR-9322427.

\vfill\eject
\centerline{\bf References}
\vskip 16 pt
\item {1.} G. Blatter, M.V. Feigel'man, V.B. Geshkenbein, A.I. Larkin,
and V.M. Vinokur, Rev. Mod. Phys. {\rm 66}, 1125 (1994).

\item {2.} W.E. Lawrence and S. Doniach, in 
{\it Proceedings of the  12$^{\rm th}$
International Conference on  Low Temperature
Physics (Kyoto, 1970)}, edited by E. Kanda
(Keigaku, Tokyo, 1971) p. 361; see also M. Tinkham, Physica C {\bf 235},
3 (1994).

\item {3.} K.B. Efetov, Zh. Eksp. Teor. Fiz. {\bf 76}, 1781 (1979)
[Sov. Phys. JETP {\bf 49}, 905 (1979)].

\item {4.} L.N. Bulaevskii, Zh. Eksp. Teor. Fiz. {\bf 64}, 2241
[Sov. Phys. JETP {\bf 37}, 1133 (1973)].

\item {5.}  S.E. Korshunov, Europhys. Lett. {\bf 11}, 757 (1990);
see also G. Carneiro, Physica C {\bf 183}, 360 (1991).

\item {6.}  B. Horovitz, Phys. Rev B {\bf 47}, 5947 (1993);
5964 (1993).

\item {7.} J.P. Rodriguez, Europhys.  Lett. {\bf 31}, 479 (1995).

\item {8.} J.P. Rodriguez, 
to be published in Phys. Rev. B
(cond-mat/9604152).
[The labels $T_*$ and $T_c$ for the transition temperatures
are interchanged in this reference in order to highlight  
the mathematical duality with the layered $XY$ model (1)].

\item {9.} M.E. Peskin, Ann. of Phys. {\bf 113}, 122 (1978).

\item {10.} Y.H. Li and S. Teitel, Phys. Rev. B {\bf 47}, 359
(1993); {\it ibid} {\bf 49}, 4136 (1994).

\item {11.} Y.M. Wan, S.E. Hebboul, D.C. Harris, and J.C. Garland,
Phys. Rev. Lett. {\bf 71}, 157 (1993); {\bf 74}, 5286 (E)
(1995).

\item {12.} Y.M. Wan, S.E. Hebboul, and J.C. Garland, Phys. Rev. Lett.
{\bf 72}, 3867 (1994).

\item {13.} B. Horovitz, Phys. Rev. Lett. {\bf 72}, 1569 (1994).

\item {14.} S.E. Korshunov and A.I. Larkin, Phys. Rev. B {\bf 46},
6395 (1992).

\item {15.} J.V. Jos\' e, L.P. Kadanoff, S. Kirkpatrick and
D.R. Nelson, Phys. Rev. B {\bf 16}, 1217 (1977).

\item {16.} L.I. Glazman and A.E. Koshelev, Phys. Rev. B {\bf 43}, 2835 (1991).

\item {17.} B.A. Huberman and S. Doniach, Phys. Rev. Lett. {\bf 43},
950 (1979); D.S. Fisher, Phys. Rev. B {\bf 22}, 1190 (1980).

\item {18.} D. Browne and B. Horovitz, Phys. Rev. Lett.
{\bf 61}, 1259 (1988); B. Horovitz, Physica B {\bf 165}, 1109
(1990).

\item {19.} I.O. Kulik, Zh. Eksp. Teor. Fiz. {\bf 51}, 1952
(1966) [Sov. Phys. JETP {\bf 24}, 1307 (1967)]; I.O. Kulik
and I.K. Yanson, {\it The Josephson Effect in Superconductive
Tunneling Structures} (Nauka, Moscow, 1970).

\item {20.} C.S. Owen and D.J. Scalapino, Phys. Rev.
{\bf 164}, 538 (1967).

\item {21.} P. Lebwohl and M.J. Stephen, Phys. Rev. {\bf 163},
376 (1967).

\item {22.}  V.M. Vinokur and A.E. Koshelev,
Zh. Eksp. Teor. Fiz. {\bf 97}, 976 (1990)
[Sov. Phys. JETP {\bf 70}, 547 (1990)].

\item {23.} J. Rubinstein, J. Math Phys. {\bf 11}, 258 (1970).

\item {24.} R.F. Dashen, B. Hasslacher, and Andr\' e Neveu,
Phys. Rev. D {\bf 11}, 3424 (1975).

\item {25.} R. Rajaraman, {\it Solitons and Instantons}
(North-Holland, Amsterdam, 1987).

\item {26.} L. Bulaevskii and J.R. Clem, Phys. Rev. B {\bf 44}, 10234 (1991).

\item {27.} A recent analysis of double-layered superconductors 
that employs
a new fermion analogy for the Lawrence-Doniach model 
finds that the cross-over shown by the magnetization
as a function of parallel magnetic field 
(see Fig. 1) is practically
destroyed by entropic pressure effects
for temperatures in the critical regime; see
J.P. Rodriguez, ``Fermion Analogy for Layered
Superconducting Films in Parallel Magnetic Field'', ICMM-CSIC report
(1995) (cond-mat/9606154).

\item {28.} A. Polyakov, Phys. Lett.  {\bf 72B}, 477 (1978);
L. Susskind, Phys. Rev. D {\bf 20}, 2610 (1979); R. Savit,
Phys. Rev. B {\bf 17}, 1340 (1978).

\item {29.} H. van Beijeren and I. Nolden, in
{\it Structure and Dynamics of Surfaces II},
edited by W. Schommers
and P. von Blanckenhagen (Springer, Heidelberg, 1987).

\item {30.} J.M. Kosterlitz, J. Phys. C{\bf 7}, 1046 (1974); see also
P. Minnhagen, Rev. Mod. Phys. {\bf 59}, 1001 (1987).

\item {31.} S. Hikami and T. Tsuneto, Prog. Theor. Phys. {\bf 63},
387 (1980); B. Chattopadhyay and S. R. Shenoy,
Phys. Rev. Lett. {\bf 72}, 400 (1994).

\item {32.} S.R. Shenoy and B. Chattopadhyay, Phys. Rev. B {\bf 51},
9129 (1995).

\item {33.} The perpendicular 2D vortices become
coherent perpendicular vortex lines that are widened by
parallel excursions of the Josephson vortices
in the low-field regime $B_{\perp}\ll B_*^{\perp}$;
see W. Janke and T. Matsui, Phys. Rev. B {\bf 42}, 10673 (1990)
and ref. 16.

\item {34.} A.M. Polyakov, Nucl. Phys. B{\bf 120}, 429 (1977);
{\it Gauge Fields and Strings} (Harwood, New York, 1987).

\item {35.} J.P. Rodriguez, Phys. Rev. B {\bf 39}, 2906 (1989);
{\it ibid.} {\bf 41}, 7326 (1990).

\item {36.} P.G. de Gennes, {\it Superconductivity of Metals and Alloys}
(Addison-Wesley, New York, 1989), chapter 3.

\item {37.}  M. Abramowitz and  I.A. Stegun,
{\it Handbook of Mathematical Functions with Formulas, Graphs, and 
Mathematical Tables} (Dover, New York, 1972); chapters 16 and 17; 
E.T. Whittaker and G.N. Watson,
{\it A Course on Modern Analysis} (Cambridge Univ. Press,
Cambridge, 1952).

\item {38.}  A.I. Larkin, Zh. Eksp. Teor. Fiz. {\rm 58}, 1466
(1970) [Sov. Phs. JETP {\bf 31}, 784 (1970)].

\item {39.} See for example S.N. Coppersmith, 
D.S. Fisher, B.I. Halperin,
P.A. Lee, and W.F. Brinkman, Phys. Rev. B {\bf 25}, 349 (1982).

\vfill\eject
\centerline{\bf Figure Captions}
\vskip 20pt
\item {Fig. 1.}   Shown is the reversible parallel magnetization of
an extreme type-II double-layer superconductor in the intermediate
regime of the mixed phase; i.e., 
the parallel fields satisfy $H_{c1}^{\parallel}\ll
H_{\parallel} \cong B_{\parallel}
\ll H_{c2}^{\parallel}$.   The cross-over field $B_*^{\parallel}$,
in particular, satisfies the latter inequalities [see Eq. (51)]. 
Also, the tail in the magnetization that appears 
at  fields beyond this scale
varies asymptotically as $H_{\parallel}^{-3}$.  
Although these results are
generally valid only for temperatures
in the Ginzburg-Landau regime,
they should provide a good lower bound for
$-4\pi M_{\parallel}$ at temperatures in the critical regime
near $T_*$ (see ref. 27).

\item {Fig. 2.} Shown is the phase diagram of a double-layer extreme type-II
superconductor in parallel external magnetic field near the
critical regime.  The KT transition temperature $T_c$ marks the
point  above which each individual layer becomes
resistive in the absence
of external magnetic field,
while $T_*$ marks the layer-decoupling transition.  
Also, the dashed line represents the
temperature profile for the parallel upper-critical field within
the Ginzburg-Landau approximation.

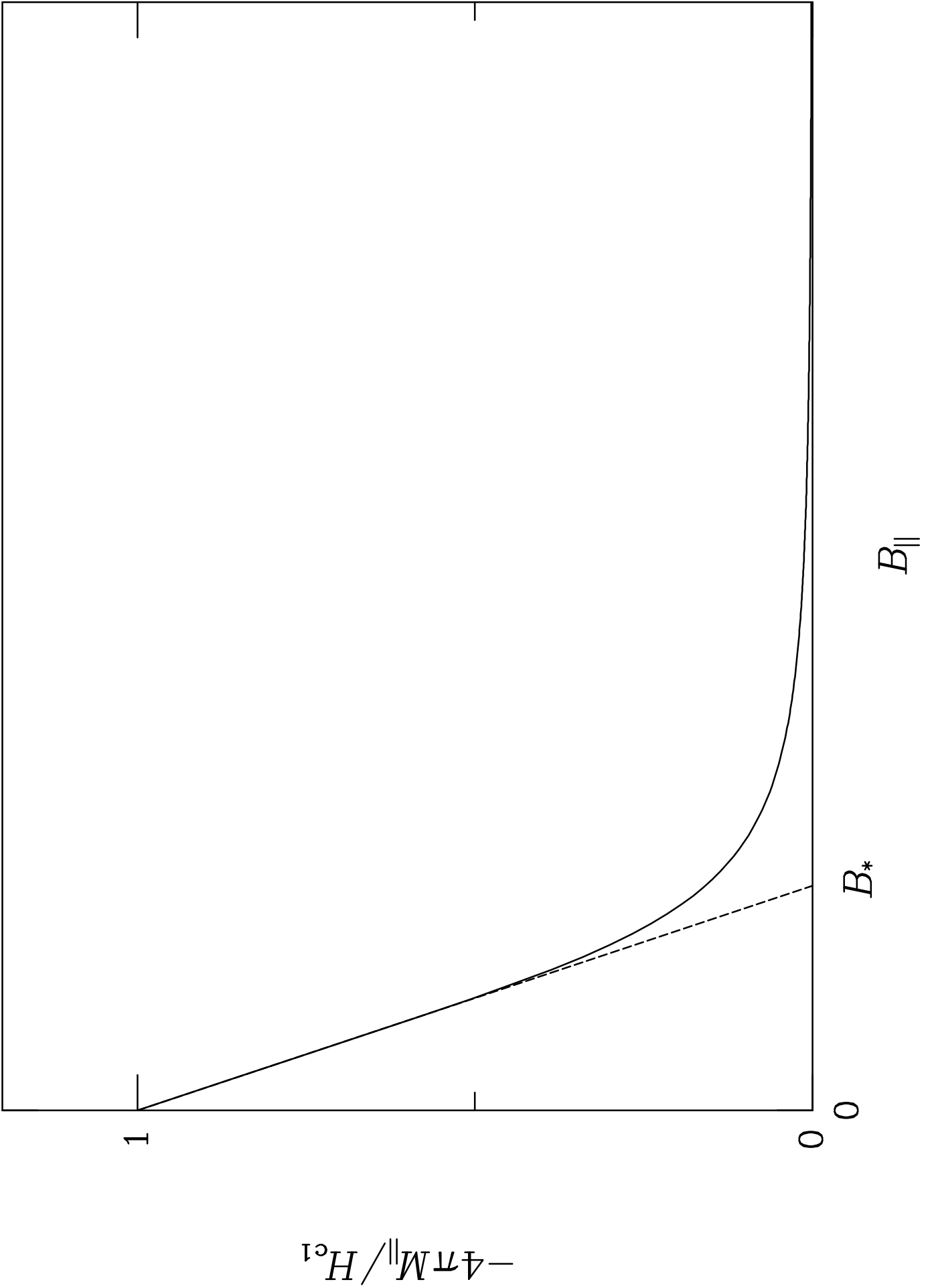
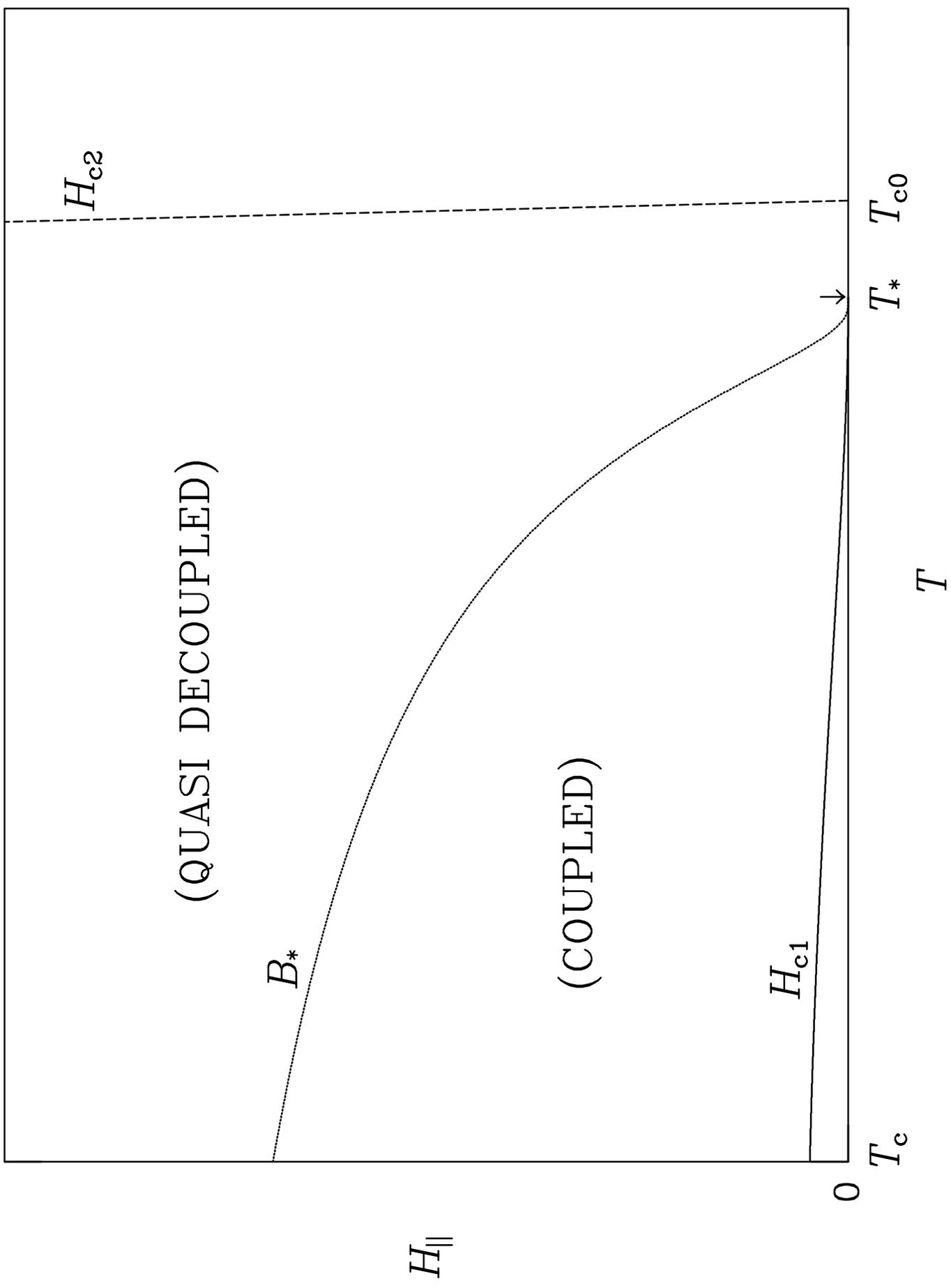
\end